\documentclass[12pt]{iopart}

\usepackage{iopams}  
\usepackage{graphicx}
\usepackage{amssymb}
\usepackage{color}
\usepackage{psfrag}
\usepackage{hyperref}

\newcommand{\bra}[2] {\langle #1 |_{#2}}
\newcommand{\ket}[2] {| #1 \rangle_{#2}}

\newcommand{\ee}[1] {\mathrm{e}^{#1}}

\newcommand{\hc}{\mathrm{h.c.}}
\newcommand{\dg}{^{\dagger}}

\begin{document}

\title[]{An excitation-dependent four-level model for quantum entanglement in photosynthetic systems}

\author{Chi-Han Chang$^{1,2}$, Agata M. Bra\'nczyk$^1$, Gregory D. Scholes$^2$ and Daniel F. V. James$^1$}

\address{$^1$ Department of Physics, University of Toronto, 60 St. George St.\\ Toronto, Ontario M5S 1A7, Canada.\\
$^2$ Department of Chemistry, University of Toronto, 80 St. George St.\\ Toronto, Ontario M5S 3H6, Canada.}
\ead{branczyk@physics.utoronto.ca}

\begin{abstract}
We model energy transfer between two coupled four-level chromophores with arbitrarily spaced energy levels. Our analysis takes into account the crucial---yet often ignored---process of initial excitation by light that is incident on the chromophores. We show that the amount of entanglement generated between the chromophores is strongly dependent on the degree of initial excitation as well as the inclusion of higher energy levels. We apply our model to the specific example of chlorophyll. Our results suggest that an excitation-dependent approach should be employed for entanglement studies on multi-level light-harvesting systems even when a two-level approximation is valid.
\end{abstract}

\maketitle

\section{Introduction}

Photosynthesis is a process utilized by organisms to produce biomass from raw materials such as carbon dioxide and water, using harvested solar energy as the driving force (Chapter 12 of \cite{campbell_biology._2005}). Central to this process are light-harvesting complexes (LHCs) which contain several light-absorbing molecules called chromophores \cite{cheng_dynamics_2009,scholes_lessons_2011,novoderezhkin_physical_2010}. Photosynthetic mechanisms are attracting increasing attention, not only because photosynthesis inspires solutions to present renewable-energy needs \cite{meyer_chemists_2011,grills_new_2010,maeda_photocatalytic_2010,mallouk_emerging_2010,vullev_biomimesis_2011,gust_mimicking_2000,gust_solar_2009}, but also due to recent evidence and predictions of non-trivial quantum coherent behaviour in certain LHCs \cite{collini_coherently_2010,engel_evidence_2007,lee_coherence_2007,turner_comparison_2011,rebentrost_role_2009,panitchayangkoon_long-lived_2010,turner_quantitative_2012,mohseni_environment-assisted_2008,pachon_physical_2011,fleming_physical_2004,fleming_quantum_2011,scholes_quantum-coherent_2010,tao_semiclassical_2010}.

Of additional interest is whether these quantum behaviours play any functional role in biology. In non-biological systems, entanglement---a quintessential quantum phenomenon---has useful application in enhanced measurements, information processing and quantum computation (see Chapter 11 of \cite{gerry_introductory_2005}, and \cite{nagata_beating_2007,giovannetti_quantum-enhanced_2004,lombardi_teleportation_2002,giovannetti_quantum_2006,gisin_quantum_2002,roos_/`designer_2006}). It is therefore intriguing to ask whether entanglement is present in biological systems and, if so, whether organisms utilize entanglement for practical purposes.

Previous entanglement studies on LHCs have employed two-level models in the single-exciton manifold \cite{ishizaki_quantum_2010,caruso_entanglement_2010,fassioli_distribution_2010,sarovar_quantum_2010}. In this picture, quantum coherence (i.e. nonzero off-diagonal elements of the density matrix in the energy eigenbasis) is a necessary and sufficient condition for entanglement \cite{hill_entanglement_1997}. An alternative model which treats chromophores as quantum harmonic oscillators (systems of an infinite number of equally spaced energy levels) excited by a coherent state of light predicts \emph{no} entanglement in the system at all \cite{tiersch_critical_2011}. It is clear that the choice of model, for the LHC and its interaction with light, plays a large role in determining the nature of the entanglement generated in the system. In this paper, we investigate the features of this role.

Inspired by known spectroscopy of chlorophylls (Chapter 1 of \cite{van_amerongen_photosynthetic_2000}, and \cite{shepanski_chlorophyll-excited_1981,renger_dissipative_1996,gradinaru_ultrafast_1998}), we consider a model with multiple arbitrarily-spaced energy levels, which we motivate in the context of previous models in Section \ref{sec:model}. In Section \ref{sec:theory}, we introduce the formalism for our multi-level model. We explicitly take into account a realistic excitation process in the preparation of the initial state, rather than assuming that the LHC absorbs single-photon pulses. We then evolve the system and calculate the entanglement, quantified by the entropy of entanglement, which is generated during energy transfer. We present numerical simulations for the entanglement in the system using realistic physical parameters in Section \ref{sec:results}, and also compare this model with simplified variations which consider fewer energy levels, before concluding in Section \ref{sec:conc}.

Before we continue, we would like to make a comment on \emph{coherence}. We understand coherence to mean ``having a phase relationship''. With this definition, it is clear that, when used in isolation, the word ``coherence'' does not mean anything---a phase relationship can only exist between two or more properties. Any confusion that arises from use of the word is likely to be related to its use in isolation. In this paper, we use the word ``coherence'' in three different contexts. The first is quantum coherence in photosynthesis, which refers to coherence (a phase relationship) between energy eigenbasis states of coupled chromophores. The second is in the context of a coherent state of light---a state of indefinite photon-number where the coherence (phase relationship) is between the different photon-number states that make up the state. The third is in terms of an analogous coherent state of a chromophore---a state with infinitely many excited levels---in which the phase relationship is between the energy eigenstates of the chromophore.

\section{Modelling light harvesting complexes}\label{sec:model}

In this section, we summarize previous LHC models and motivate our model.

\subsection{Two-level system in the single-exciton manifold}

By far the most common model for a chromophore is that of a two-level system, consisting of an electronic ground state $\ket{0}{}$ and an excited state $\ket{1}{}$ \cite{ishizaki_quantum_2010,caruso_entanglement_2010,fassioli_distribution_2010,sarovar_quantum_2010}. Organisms containing these chromophores live at low light conditions under which excitation of the chromophore to a higher excited state is unlikely. This is used to justify the two-level approximation, which states that the system will only contain at most a single excitation; as well as the single-exciton assumption, which states that the system will contain \emph{only} a single excitation. In the single-excitation manifold, the initial excited state of an LHC is assumed to have one chromophore in the $\ket{1}{}$ state and all others in $\ket{0}{}$. These models have predicted long lived entanglement in a chlorophyll-containing complex \cite{ishizaki_quantum_2010} and the Fenna-Mathews-Olson (FMO) complex \cite{caruso_entanglement_2010,fassioli_distribution_2010,sarovar_quantum_2010}.

However, implicit in the preparation of this initial state is a light-matter interaction between a two-level system and a \emph{single} photon---a highly non-classical state of light. This single photon must be absorbed with unit probability to generate a state with exactly one excitation, i.e.  $\ket{1}{}$.

We are at present unaware of any biological mechanism or experiment resulting in the absorption of exactly one photon by a chromophore. After interaction with classical light like a laser or sunlight, a chromophore would be excited into a superposition or statistical mixture of $\ket{0}{}$ and $\ket{1}{}$ respectively (Chapter 4 of \cite{gerry_introductory_2005}). This can be thought of as allowing for the possibility that light is not absorbed, leaving the system with a ground state population. In fact, only around $1\%$ of input sunlight is estimated to be absorbed by chlorophyll under natural conditions (see Chapter 5 of \cite{blankenship_molecular_2002}). Accordingly, the single excitation assumption imposes an excited state that may not correspond to a realistic situation. We note that even if it were a single photon that was  incident on the chromophore, the chromophore would still be excited into a superposition of $\ket{0}{}$ and $\ket{1}{}$ unless one could guarantee that the photon were absorbed. 

\subsection{Quantum harmonic oscillator model}

An alternative model which treats the chromophore as a quantum harmonic oscillator (QHO) was introduced by Tiersch \emph{et al.} \cite{tiersch_critical_2011}. Such a system consists of an infinite number of equally spaced energy levels. Under low-light conditions, the extra levels may seem superfluous and the two-level model should suffice, however the rationale for the QHO model becomes apparent when considering the light-matter interaction.

In the dipole and rotating wave approximations, the evolution of the system due to the light-matter interaction between a coherent state of light and the QHO is given by the displacement operator $\hat D(\alpha)=\exp(\alpha\hat a\dg-\alpha^*\hat a)$. This immediately gives the analogy between the state prepared by this light-matter interaction and the coherent state $\ket{\alpha}{}=\hat D(\alpha)\ket{0}{}$. An analogy is also made between the FMO complex and a multi-armed interferometer. The propagation of a single excitation through the complex is analogous to the propagation of a single photon through the interferometer---both will clearly lead to entanglement \cite{van_enk_single-particle_2005}. In contrast, the propagation of the coherent state of the chromophore through the complex is analogous to the propagation of a coherent state of light through the interferometer, which results in \emph{no} entanglement being created.

Tiersch \emph{et al.} show that higher-level truncation of this unentangled state introduces the perception of entanglement. A more dramatic increase in the perceived entanglement is also observed when the ground state of the coherent state is disregarded, which corresponds to working in the single-excitation manifold.

Although this model realistically considers the process of initial state preparation, it only deals with a very special quantum mechanical state and the results may not be applicable to systems that deviate from this unique case. In particular, the energy level structure of a QHO does not represent that of a chromophore very well.

\subsection{Mutli-level model}

The two models described in the previous sections lead to contradictory results. This indicates that initial state preparation and the details of the level structure should be taken into account carefully.  In our model, we consider a realistic light-matter interaction for the state preparation as well as a more complete description of the electronic structure of the system itself.

Certain chromophores do possess higher excitation levels that cannot be neglected under typical light conditions. The coherent state of a QHO is a convenient model for a many-level system, but it restricts the system to a ladder of equally spaced energy levels. The chromophores in question do not have equal energy spacing nor are they necessarily in the form of a ladder system.

A classic example of a multi-level system is chlorophyll in higher plants. The absorption spectrum of chlorophyll has two intense bands denoted as Q$_y$ and Soret, corresponding to electronic transitions from ground to two separate excited states (Chapter 1 of \cite{van_amerongen_photosynthetic_2000}). Another band denoted as Q$_x$ is weaker, and thus will not be considered in our calculation. Excited state absorption has also been observed experimentally in these systems
 \cite{dage_density_1999,renger_dissipative_1996,gradinaru_ultrafast_1998}. Another example of multi-level systems are carotenoids: chromophores that have at least three observed transitions, i.e. the $S_0\leftrightarrow S_2$ transition, the $S_0\leftrightarrow S_n$ transition and the two-photon allowed $S_0\leftrightarrow S_1$ transition (Chapter 1 of \cite{van_amerongen_photosynthetic_2000}).

\begin{figure}[h]
\begin{center}
\includegraphics[width=0.9\columnwidth]{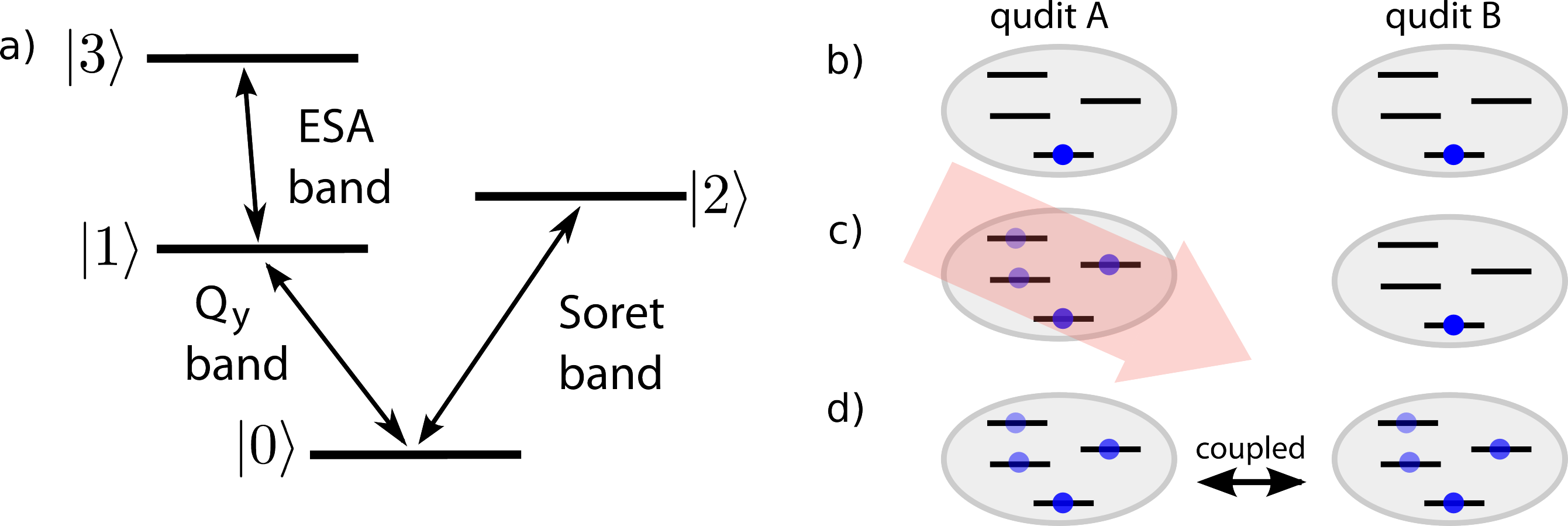}
\caption{a) Electronic structure of the four-level system (qudit). Transitions from $\ket{0}{}$ to $\ket{1}{}$ and from $\ket{0}{}$ to $\ket{2}{}$ are analogous to transitions leading to the Q$_y$ and Soret absorption bands in chlorophyl, respectively \cite{van_amerongen_photosynthetic_2000,shepanski_chlorophyll-excited_1981}. The transition from $\ket{1}{}$ to $\ket{3}{}$ describes excited state absorption (ESA) also witnessed in experiments \cite{dage_density_1999,renger_dissipative_1996,gradinaru_ultrafast_1998}. b) Two identical uncoupled qudits begin in the ground state. c) qudit A is excited by classical light (refer to Section \ref{subsec:ExProcess}). d) energy transfer occurs between qudit A and B (refer to Section \ref{sec:eet}). }
\label{fig:qudit}
\end{center}
\end{figure}

With these systems in mind, we develop a more general multi-level model. We consider a four-level system with arbitrarily-spaced energy levels, depicted in Figure \ref{fig:qudit} a). In the language of quantum information \cite{nielsen_quantum_2011}, this system is referred to as a \emph{qudit} (a $d$-level generalization of the quantum bit, or \emph{qubit}). To model energy transfer, we take the simplest case of a coupled dimer consisting of two qudits. The state is prepared by interacting a qudit with a pulse of coherent classical light. We note that although excitation may not actually occur in the site (chromophore) basis but rather in the exciton basis, we will consider the former case in order to put our work in the context of previous work on the subject which has considered site-basis excitation \cite{tiersch_critical_2011,ishizaki_quantum_2010,caruso_entanglement_2010,fassioli_distribution_2010,sarovar_quantum_2010}. 

In this paper, we do not consider any decoherence mechanisms which arise from the interaction of the chromophores with their environment, nor do we consider line broadening or static disorder.  These effects are critical for a complete description of photosynthetic systems, however, they are beyond the scope of this paper. Our model is idealized in this sense and instead, we concentrate on the consequences of different excitation processes and energy level structures. 

\section{Theory}\label{sec:theory}

To model the entanglement generated during energy transfer in a multi-chromophoric system, we study the simplest case of two coupled qudits. We prepare the initial state by considering the light-matter interaction between classical light and one qudit, and then calculate the evolution of the entanglement between the two coupled qudits.

\subsection{Initial excitation}\label{subsec:ExProcess}

For excitation of chromophore A in the site basis, we first assume that the chromophores are uncoupled. We take the total Hamiltonian for the qudit-light system to be
\begin{equation}
\hat{H}_{1}=\hat{H}_{\mathrm{d}}+\hat{H}_{\mathrm{lm}}\;,
\end{equation}
where $\hat{H}_{\mathrm{d}}$ is the Hamiltonian of the qudit given by
\begin{equation}
\hat{H}_{\mathrm{d}}=\sum_{n=0}^{3}\hbar\omega_{n}\ket{n}{}\bra{n}{},\label{eq:Hm}
\end{equation}
where $\hbar\omega_{0}$ is the ground state energy and $\hbar\omega_{1-3}$ are energies of the three excited states.

As the size of a chlorophyll molecule is roughly 1nm---much shorter than the wavelength of visible light---the dipole approximation for absorption can be invoked (Chapter 3.6 of \cite{craig_molecular_1998}), giving the light-matter interaction Hamiltonian
\begin{equation}
\hat{H}_{\mathrm{lm}}=-\vec{E}(t)\cdot\hat{d},
\end{equation}
where $\hat{d}$ is the dipole moment operator and $\vec{E}=\vec{E}^{(+)}(t)+\vec{E}^{(-)}(t)$ is a classical electric field given by
\begin{equation}
\vec{E}^{(+)}(t)=E_{0}\ee{i\omega_{\mathrm{L}}t}\vec{\mathbf{e}},
\end{equation}
where $E_0$ is the electric field amplitude, $\omega_{\mathrm{L}}$ is the carrier frequency and $\vec{\mathbf{e}}$ is the polarization vector. We can define a co-rotating frame, where the transformation is given by the unitary operator $\hat{U}$ such that
\begin{equation}
\ket{\phi_{\mathrm{cr}}}{}\:=\hat{U}\,\ket{\phi}{}\,,
\end{equation}
where $\ket{\phi}{}$ is the state in the Schr\"odinger picture. We aim to find a time-independent Hamiltonian such that the dynamics of the system is given by (Chapter 9 of \cite{tannor_introduction_2007})
\begin{equation}
\ket{\phi_{\mathrm{cr}}(t)}{}\:=\ee{-\frac{i}{\hbar}\hat{H}_{\mathrm{cr}}t}\:\ket{\phi_{\mathrm{cr}}(0)}{},
\end{equation}
where 
\begin{equation}
\hat{H}_{\mathrm{cr}}={}\hat{U}^{\dagger}\hat{H}_{1}(t)\hat{U}-i\hbar \hat{U}^{\dagger}\frac{\partial}{\partial t}\hat{U}\,.
\end{equation}

We find that the operator
\begin{equation}\label{eq:U}
\hat{U}={}\left(\begin{array}{cccc}\ee{i t\omega_L} & 0 & 0 & 0 \\0 & 1 & 0 & 0 \\0 & 0 & 1 & 0 \\0 & 0 & 0 & \ee{-i t\omega_L}\end{array}\right)\,,
\end{equation}
gives the following time-independent Hamiltonian in the rotating wave approximation
\begin{eqnarray}\label{eq:hcr}
\hat{H}_{\mathrm{cr}} ={} &\left(
\begin{array}{cccc}
\hbar \omega_{\mathrm{L}} & E_0 d_{01} & E_0
d_{02} & 0 \\
E_0 d_{10} & \hbar \omega _1 & 0 & E_0
d_{13} \\
E_0 d_{20} & 0 & \hbar \omega _2 & 0 \\
0 & E_0 d_{31} & 0 & \hbar \omega
_3-\hbar \omega_{\mathrm{L}}
\end{array}
\right)\,,
\end{eqnarray}
where ${d}_{nm}=\vec{\mathbf{e}}\cdot\bra{n}{}\hat{d}\ket{m}{}$ and we have made use of the experimental observation that in chlorophyll, the most physically relevant transitions are those illustrated in Figure \ref{fig:qudit} a).

\subsection{Inter-qudit excitation energy transfer (EET)}\label{sec:eet}

To model energy transfer, we consider two coupled qudits. We take the Hamiltonian for this system to be
\begin{equation}\label{eq:H2}
\hat{H}_{2}=\hat{H}_{\mathrm{dd}}+\hat{H}_{\mathrm{c}}\;,
\end{equation}
where $\hat{H}_{\mathrm{dd}}=\hat{H}_{\mathrm{d}}\otimes \mathbf{I}_{\mathrm{B}}+\mathbf{I}_{\mathrm{A}}\otimes \hat{H}_{\mathrm{d}}$ is the Hamiltonian for the free qudits, where $\hat{H}_{\mathrm{d}}$ is defined in (\ref{eq:Hm}), and
\begin{eqnarray}\nonumber
\hat{H}_{\mathrm{c}} ={} &\Big(J_{10,01}\ket{10}{}\bra{01}{}+J_{20,02}\ket{20}{}\bra{02}{}+J_{13,31}\ket{13}{}\bra{31}{} \\
& +J_{20,01}\ket{20}{}\bra{01}{}+J_{10,02}\ket{10}{}\bra{02}{}+J_{12,30}\ket{12}{}\bra{30}{} \\\nonumber
& +J_{11,30}\ket{11}{}\bra{30}{}+J_{11,03}\ket{11}{}\bra{03}{}\Big)+\hc
\label{eg:coupH}
\end{eqnarray}

This coupling Hamiltonian is constructed according to the observation that only transitions between $\ket{0}{}$ and $\ket{1}{}$; $\ket{0}{}$ and $\ket{2}{}$; and $\ket{1}{}$ and $\ket{3}{}$ can occur in the systems under consideration \cite{van_amerongen_photosynthetic_2000,renger_dissipative_1996,gradinaru_ultrafast_1998,dage_density_1999}. 

\subsection{The final state}\label{sec:final}

To calculate the final state, we begin with both uncoupled chromophores in the ground state $\ket{\psi(0)}{\mathrm{a,b}}=\ket{0}{\mathrm{a}}\ket{0}{\mathrm{b}}$ as indicated in Figure \ref{fig:qudit} b). After interaction of qudit $A$ with light, as indicated in Figure \ref{fig:qudit} c), the state is 
\begin{eqnarray}
\ket{\psi_{\mathrm{in}}(T)}{\mathrm{a}}=\hat{U}\dg\ee{-\frac{i}{\hbar}\hat{H}_{\mathrm{cr}} T}\hat{U}\ket{0}{\mathrm{a}}\,,
\end{eqnarray}
where  $\hat{U}$ is defined in (\ref{eq:U}) and $\hat{H}_{\mathrm{cr}} $ is given in (\ref{eq:hcr}). We take $\ket{\psi_{\mathrm{in}}(T)}{\mathrm{a,b}}=\ket{\psi_{\mathrm{in}}(T)}{\mathrm{a}}\ket{0}{\mathrm{b}}$ to be the initial state in terms of the impending energy transfer between the coupled qudits. We now let the qudits couple to each other to allow for the energy transfer, as indicated in Figure \ref{fig:qudit} d). The final state is
\begin{eqnarray}
\ket{\psi(t)}{\mathrm{a,b}}&={}&\ee{-\frac{i}{\hbar}\hat{H}_{2}t}\ket{\psi_{\mathrm{in}}(T)}{\mathrm{a,b}}\,,
\end{eqnarray}
where $t\approx t+T$ since $t\gg T$, $T$ is the duration of the light-matter interaction, and $\hat{H}_{2}$ is defined in (\ref{eq:H2}).
\subsection{Measure of entanglement}

Any pure bipartite state
\begin{equation}
\ket{\psi_{\mathrm{s}}(t)}{}\:=\sum_{m=0}^{3}\sum_{n=0}^{3}c_{mn}\ket{m}{}\ket{n}{},
\end{equation}
can be decomposed into what is known as the Schmidt-decomposed form
\begin{equation}
\ket{\psi_{\mathrm{s}}(t)}{}\:=\sum_{k=0}^{3}s_{k}\ket{\alpha_{k}}{}\ket{\beta_{k}}{},
\end{equation}
where $\{\ket{\alpha_{k}}{}\}$ and $\{\ket{\beta_{k}}{}\}$ form orthonormal bases and the diagonal matrix $s$ is found by performing a singular value decomposition of the matrix $c$ (Section 20.2 of \cite{meystre_elements_2007}).

We use the entropy of entanglement, defined as the von Neumann entropy of one of the reduced states of a bipartite system, as the measure of the entanglement over time for the two-qudit system. In terms of the Schmidt values $s_k$, this is \cite{meystre_elements_2007,wootters_entanglement_1998}
\begin{equation}
E\:[\ket{\psi_{\mathrm{s}}(t)}{}]\;=-\sum_{k=0\,}^{3}s_{k}^{2}\, \log_{2}s_{k}^{2}.
\end{equation}

The entropy of entanglement ranges from zero for a product state to $\log_2N$ for a maximally entangled state of two $N$-state particles, which in our case is $\log_24=2$. 

\subsection{Parameters}\label{subsec:parameter}

We estimate reasonable physical parameters by considering an analogous photosynthetic system of two weakly-coupled pairs of strongly-coupled qubits \cite{scholes_adapting_2001} and using numerical values from \cite{mirkovic_ultrafast_2007}. For a detailed calculation, refer to  \ref{sec:appdxPara}. Our estimated parameters are summarized in Table \ref{tab:para}.

\begin{table}[h!]
\begin{center}
\begin{tabular}{lll}
\hline
\hline
Level Energy~ & Dipole Strength~ & Coupling\\
\hline
\hline
$\omega_0=0$&&\\
$\omega_1=\omega$ &$|\vec{d}_{10}|=d$ &$J_{10,01}=J$ \\
$\omega_2=1.04 \omega$ &$|\vec{d}_{20}|=0.94d$&$J_{20,02}=0.50J$\\
& &$J_{20,01}=-0.67 J$\\
& &$J_{10,02}=0.72 J$\\
$\omega_3=2 \omega$ &$|\vec{d}_{31}|=d$ &$J_{13,31}=0.90J$ *\\
& &$J_{11,30}=J_{11,03}=0.81J$ *\\
& &$J_{12,30}=0.76J$ *\\
\hline
\hline
\end{tabular}
\end{center}
\caption{Estimated numerical values for relative energy spacings and couplings. * denotes values that are empirically assigned by assuming that $J$ is slightly larger than $J_{13,31}$, which is slightly larger than $J_{11,30}=J_{11,03}$, which is in turn larger than $J_{12,30}$. All other values are estimated according to \cite{scholes_adapting_2001}, \cite{mirkovic_ultrafast_2007} and \cite{collini_coherently_2010}. Typical values are $\omega=2.99\times10^{15}\mathrm{s}^{-1}$, $d=5\mathrm{D}$ and $J=-1.25\times 10^{12}\mathrm{J}$ (see \ref{sec:appdxPara}).}
\label{tab:para}
\end{table}

To facilitate interpretation of the numerical results, we introduce four convenient dimensionless parameters, given in Table \ref{tab:dimParac}.
\begin{table}[h!]
\begin{center}
\begin{tabular}{ll}
\hline
\hline
Parameter& Interpretation\\
\hline
\hline
$\gamma=E_0 d T/ \hbar$&degree of initial excitation\\
$\delta=\omega T=\gamma\omega\hbar/E_0d$ &\\
$\gamma_2=J t/ \hbar$ &coupling strength and/or evolution time\\
$\delta_2=\omega t=\gamma_2\omega\hbar/J$ &\\
\hline
\hline
\end{tabular}
\end{center}
\caption{Summary of dimensionless parameters. $E_0$ is the electric field amplitude, $d$ is related to the dipole moments (see Table \ref{tab:para}), $T$ is the light-matter interaction time, $\omega$ is related to the level energies (see Table \ref{tab:para}), $J$ is related to the coupling strengths (see Table \ref{tab:para}) and $t$ is the evolution time, where $t\gg T$.  We study how entanglement changes as $\gamma$ and $\gamma_2$ vary while $\delta$ is kept fixed. Note that $\delta_2$ varies as a function of $\gamma_2$ as $\delta_2=\gamma_2 w/J$. A typical value is $\gamma=0.41$ (see \ref{sec:appdxLight}).}
\label{tab:dimParac}
\end{table}

\newpage

\section{Results and discussion}\label{sec:results}

In this section, we numerically investigate the entanglement between the two chromophores as a function of the degree of excitation and the dynamics of the state.

\subsection{Excitation dependence}\label{subsec:exDepend}

\begin{figure}[h]
\begin{center}
\includegraphics[width=0.9\columnwidth]{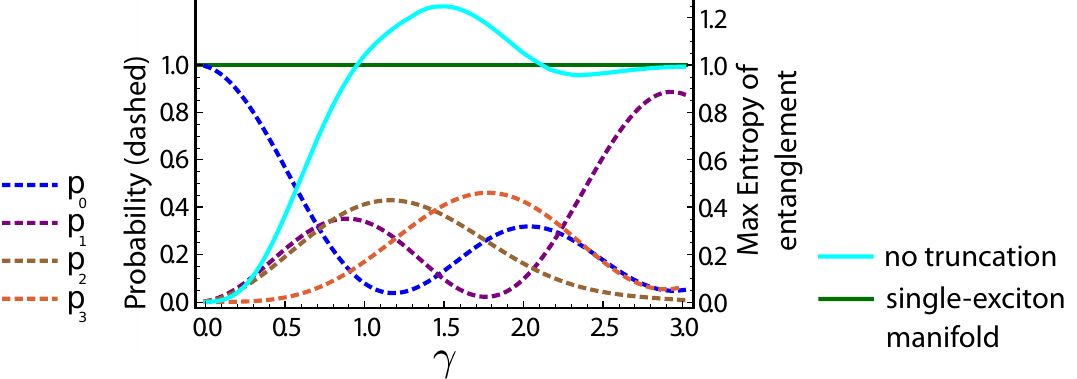}
\end{center}
\caption{Probability $p_n=|\bra{\psi_{\mathrm{in}}(T)}{\mathrm{a}}\ket{n}{\mathrm{a}}|^2$ for qudit A and the resultant maximum entanglement $\mathrm{max}(E[\ket{\psi(t)}{\mathrm{a,b}}])$ after inter-qudit coupling. Note that, in accordance with the definitions in Table \ref{tab:dimParac}, a more intense light field leads to a larger $\gamma$. Within $0<\gamma<1$ which is close to the typical experimental condition of $\gamma=0.41$, increasing $\gamma$ gives a higher probability to be in the excited states and subsequently a larger maximum entanglement between qudits.}
\label{fig:fig2}
\end{figure}

To investigate the excitation dependence on the prepared initial state, we plot probabilities $p_n=|\bra{\psi_{\mathrm{in}}(T)}{\mathrm{a}}\ket{n}{\mathrm{a}}|^2$ as a function of $\gamma$ in Figure \ref{fig:fig2}. The maximum entanglement between the two chromophores $E_{\mathrm{max}}=\mathrm{max}(E[\ket{\psi(t)}{\mathrm{a,b}}])$ as a function of $\gamma$ is also shown  in Figure \ref{fig:fig2}. We find that $E_{\mathrm{max}}$ is strongly dependent on the degree of the initial excitation. 

\newpage

For small values of $\gamma$, e.g. those corresponding to typical light conditions such as $\gamma\approx 0.41$ (as calculated in \ref{sec:appdxLight}), $E_{\mathrm{max}}$ increases with the level of initial excitation. Comparison of this with the maximum entropy of entanglement when assuming a single-exciton manifold, which is equivalent to assuming an initial state of $\ket{\psi_{\mathrm{in}}(T)}{\mathrm{ab}}=\ket{10}{\mathrm{ab}}$, is also shown in Figure \ref{fig:fig2}.

Incidentally, at $\gamma=3$, qudit A has a probability  of  $p_1\approx 0.9$. When qudit A couples to qudit B, the $\ket{2}{}$ and $\ket{3}{}$ contributions in either qudit remain low and entanglement can reach $E_{\mathrm{max}}\approx 1$, also shown in Figure \ref{fig:fig2}. This special regime resembles the case of the single-excitation manifold model. In the other limit when $\gamma\rightarrow0$, the entanglement vanishes, just as it does in the QHO model \cite{tiersch_critical_2011}. These distinct entanglement outcomes depict the dependence of entanglement on light-matter interaction.

\begin{figure}[h]
\begin{center}
\includegraphics[width=0.6\columnwidth]{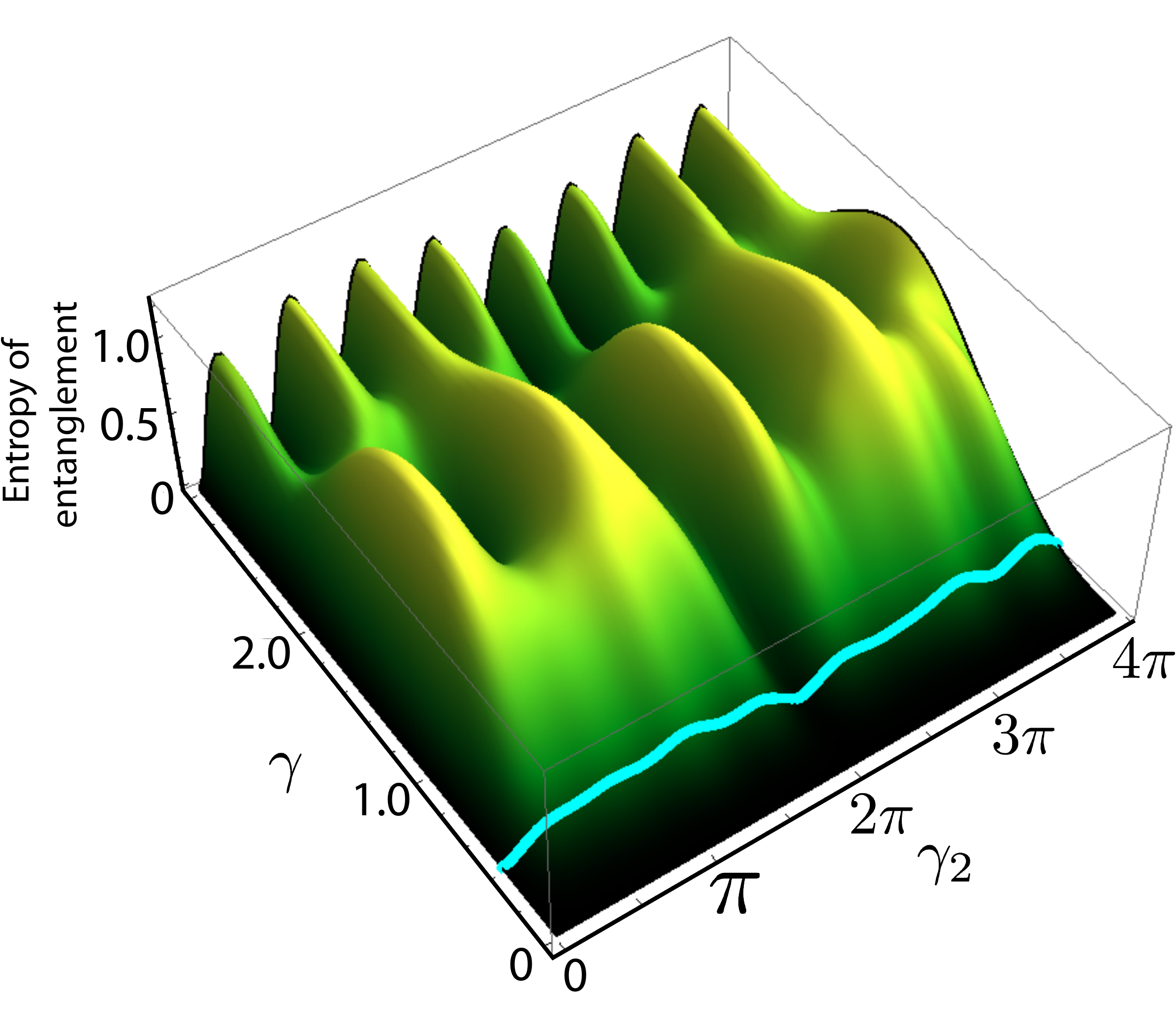}
\end{center}
\caption{Dependence of the entropy of entanglement on the degree of initial excitation. In the regime of $0<\gamma<1$, a more highly-excited qudit generates a larger entropy of entanglement when it is coupled to another ground-state qudit. The cross section at $\gamma=0.41$ is plotted in Figure \ref{fig:Ent} in cyan. Clearly, the degree of initial excitation quantified by $\gamma$ largely determines the entanglement profile, and should thus be considered when examining entanglement in LHCs.}
\label{fig:3D}
\end{figure}

Figure \ref{fig:3D} illustrates the dramatic dependence of the entropy of entanglement on $\gamma_2$ as well as $\gamma$. For small values of $\gamma$ the entropy of entanglement increases with the level of initial excitation for all values of $\gamma_2$. 

\begin{figure}[h]
\begin{center}
\includegraphics[width=0.8\columnwidth]{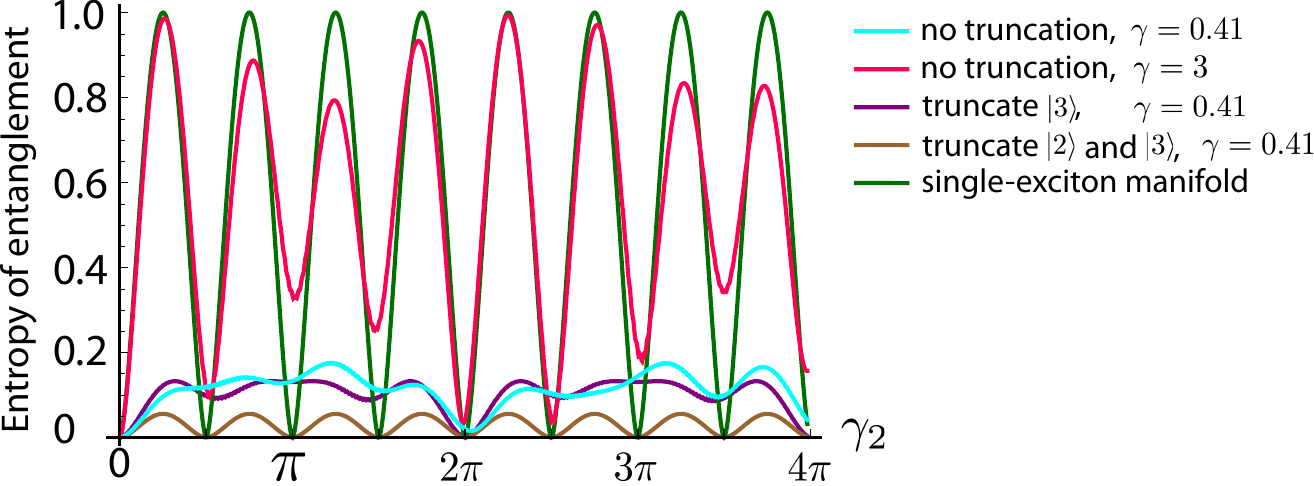}
\end{center}
\caption{The entropy of entanglement for the four-level system for two different values of $\gamma$; the three-level and two-level simplifications; as well as the case of the single-exciton manifold.  }
\label{fig:Ent}
\end{figure}

\subsection{Effect of neglecting higher levels}

To evaluate the effectiveness of simplified versions of our model, we consider a number of truncations of the four-level system. The Hamiltonians have been modified accordingly to incorporate the simplification. The entropy of entanglement as a function of $\gamma_2$, for each truncated system compared with the untruncated case, is shown in Figure \ref{fig:Ent}. The case of the single-excitation manifold is also shown. Neglecting the highest excited state (i.e. the $\ket{3}{}$ state) does not significantly change the entanglement profile compared to the four-level case, as can be expected from the fact that the $\ket{3}{}$ state already has minimal contribution even before being neglected. Accordingly, we consider the three-level simplification to be a good approximation.

Subsequent neglection of the $\ket{2}{}$ level reduces the maximum perceived entanglement by approximately half. A two-level simplification may therefore provide a lower bound for the estimate of entanglement for a pair of four-level systems with a V-like configuration. This is in contrast with the QHO model (a ladder system) where neglecting higher levels always increases the perceived entanglement \cite{tiersch_critical_2011}, emphasizing the importance of careful consideration of the energy level structure of the system. Notice that even in the case of a two-level approximation that considers the light-matter interaction, the entanglement profile is dramatically different to that of the single-exciton manifold, showing that one does not imply the other.

It is perhaps surprising that neglecting the possibility of Soret absorption has a much more dramatic effect on the entanglement than does neglecting the possibility of excited-state absorption.

\section{Conclusion}\label{sec:conc}

We provide a general model for a pair of four-level systems in a V-like configuration with arbitrary energy spacings, and input numerical parameters to obtain physically reasonable values for the entropy of entanglement.

We have demonstrated that the entanglement profile largely depends on the degree of initial excitation, which incorporates the light intensity, field duration and transition dipoles. Within a realistic range of physical parameters, a more excited chromophore generally produces larger entanglement upon interaction with the other initially un-excited chromophore. Under extreme degrees of initial excitation, our model can produce a maximum entanglement either as high as in a two-level model in the single-exciton manifold or as low as in a harmonic oscillator model. This dramatic difference demonstrates the importance of considering the light-matter interaction.

Surprisingly, if a system has four levels in this particular configuration, as in chlorophylls, the two-level simplification may provide a lower estimate of entanglement. Nevertheless, the excitation process should be considered to determine the initial state. Otherwise, the single-exciton manifold may significantly overestimate the entanglement.

We did not consider any decoherence mechanisms which arise from the interaction of chromophores with their environment, nor did we consider line broadening or static disorder. It will be interesting to incorporate our findings into a decoherence model to study the entanglement in a photosynthetic complex like chlorophyll---one of the most common yet sophisticated systems in nature.

\section{Acknowledgments}

The authors would like to thank Daniel B. Turner and Francesca Fassioli for helpful discussions, and acknowledge DARPA (QuBE) and NSERC for financial support.

\appendix

\section{Parameter estimation}\label{sec:appdxPara}

For chlorophylls, numerical values of certain couplings, such as $J_{20,02}$ in (\ref{eg:coupH}), are not available. Thus we seek estimates of parameters that are within the physical order of magnitude for photosynthetic systems. It has been shown in \cite{scholes_adapting_2001} that a system consisting of four qubits with the following properties---qubits A$_1$ and A$_2$ are strongly coupled, qubits B$_1$ and B$_2$ are strongly coupled, while qubits A$_i$ are weakly coupled to qubits B$_i$---can be treated as a two-qudit system. 

These conditions are satisfied by the photosynthetic protein Phycocyanin 645, where we define the chromophores MBV 19 A, PCB 158 C, MBV 19 B, and PCB 158 D as qubits A$_1$, A$_2$, B$_1$ and B$_2$, respectively (see Table S.1 in \cite{huo_theoretical_2011}). Chromophores are often referred to as \emph{sites} in the literature. These four sites effectively form a two-qudit system: sites A$_1$ and A$_2$ couple to form qudit A while sites B$_1$ and B$_2$ couple to form qudit B. We split the Frenkel exciton Hamiltonian (which is given in the single-exciton subspace) for the combined four-site system into two parts:
\begin{eqnarray}
\hat{H}_{\mathrm{AB}}=\left(\begin{array}{cc}\hat{H}_{A} &0 \\0 & \hat{H}_{B}\end{array}\right)\ &+ &\hat{V},
\end{eqnarray}
$\hat{V}$ incorporates the weak couplings between $A_i$ and $B_i$, and is given in (\ref{eq:V}). $\hat{H}_{A}$ and $\hat{H}_{B}$ are the Frenkel exciton Hamiltonians for the combined $A_1$ and $A_2$ system and the combined $B_1$ and $B_2$ system respectively:
\begin{eqnarray}
\hat{H}_{A}&=&\left(\begin{array}{cc}e_{A_1} & v_{A_1,A_2} \\v_{A_2,A_1} & e_{A_2}\end{array}\right)\,,\\
\hat{H}_{B}&=&\left(\begin{array}{cc}e_{B_1} & v_{B_1,B_2} \\v_{B_2,B_1} & e_{B_2}\end{array}\right)\,.
\end{eqnarray}

Diagonalizing $\hat{H}_{A}$ and $\hat{H}_{B}$ gives eigenvectors
\begin{eqnarray}\label{eq:alpha}
\ket{1}{A}&=& a_{11}\ket{e_{A_1}}{}+a_{12}\ket{e_{A_2}}{}\,,\\
\ket{2}{A}&=& a_{21}\ket{e_{A_1}}{}+a_{22}\ket{e_{A_2}}{}\,,
\end{eqnarray} 
and
\begin{eqnarray}
\ket{1}{B}&=& b_{11}\ket{e_{B_1}}{}+b_{12}\ket{e_{B_2}}{}\,,\\\label{eq:beta}
\ket{2}{B}&=& b_{21}\ket{e_{B_1}}{}+b_{22}\ket{e_{B_2}}{}\,,
\end{eqnarray} 
with corresponding eigenvalues $\omega_{A,1}$, $\omega_{A,2}$ and $\omega_{B,1}$, $\omega_{B,2}$ respectively, i.e. the energies of the two lower excited states of each qudit.

Energies of the third excited states $\ket{3}{A}$ and $\ket{3}{B}$ are estimated to be $\omega_{A,3}=2\omega_{A,1}$ and $\omega_{B,3}=2\omega_{B,1}$ based on a typical chlorophyll molecule where the energy of the $\ket{1}{}\rightarrow\ket{3}{}$ transition (excited state absorption) is close to the energy of the $\ket{0}{}\rightarrow\ket{1}{}$ transition (Q$_y$ absorption) \cite{dage_density_1999,renger_dissipative_1996,gradinaru_ultrafast_1998}. For the ground states $\ket{0}{A}$ and $\ket{0}{B}$, we define $\omega_0\equiv\omega_{A,0}=\omega_{B,0}=0$.

Starting from the values
\begin{eqnarray}
\hat{H}_{A}&=&\left(\begin{array}{cc}16050 & -87\\
-87 & 15808\end{array}\right)\,
\end{eqnarray}
and
\begin{eqnarray}
\hat{H}_{B}&=&\left(\begin{array}{cc} 16373 & 86 \\
86 & 15889\end{array}\right)\,,
\end{eqnarray}
taken from \cite{huo_theoretical_2011}, we find that $w_{A,1}\approx w_{B,1}$ and $w_{A,2}\approx w_{B,2}$. Thus for simplicity we take
\begin{eqnarray}
\omega_{1}&=&\omega_{A,1}=\omega_{B,1}=2.99\times10^{15} \mathrm{s}^{-1}\,,\\
\omega_{2}&=&\omega_{A,2}=\omega_{B,2}=3.11\times10^{15} \mathrm{s}^{-1}\,.
\end{eqnarray}

Couplings between the qudits are given by
\begin{eqnarray}\label{eq:coup}
J_{10,01}&=&\bra{1}{A}\hat{V}\ket{1}{B}\,,\\
J_{20,02}&=&\bra{2}{A}\hat{V}\ket{2}{B}\,,\\
J_{10,02}&=&\bra{1}{A}\hat{V}\ket{2}{B}\,,\\
J_{20,01}&=&\bra{2}{A}\hat{V}\ket{1}{B}\,,
\end{eqnarray}
where the states are defined in (\ref{eq:alpha}-\ref{eq:beta}) and 
\begin{eqnarray}\label{eq:V}
\hat{V} &={} &\left(
\begin{array}{cccc}
0 & 0 & 4 & -3\\
0 & 0 & 3 & 8\\
4 & 3 & 0 & 0\\
-3 & 8 & 0 & 0\\
\end{array}
\right)\mathrm{cm}^{-1}\,
\end{eqnarray}
is based on parameters given in \cite{huo_theoretical_2011}. This gives $J\equiv J_{10,01}=-1.25\times10^{12} \mathrm{J}$ and other couplings are given as ratios in Table \ref{tab:para}.

The transition dipoles for qudit A are given by \cite{scholes_adapting_2001}
\begin{eqnarray}
\vec{d}_{A,10}&=&\bra{1}{A}\hat{d}\ket{0}{A}\,,\\
\vec{d}_{A,20}&=&\bra{2}{A}\hat{d}\ket{0}{A}\,,\\
\vec{d}_{B,10}&=&\bra{1}{B}\hat{d}\ket{0}{B}\,,\\
\vec{d}_{B,20}&=&\bra{2}{B}\hat{d}\ket{0}{B}\,,
\end{eqnarray}
where the states are defined in (\ref{eq:alpha}-\ref{eq:beta}) and 
\begin{eqnarray}
\bra{e_{A_1}}{}\hat{d}\ket{0}{}&=&(-1.42 , 4.54 , -13.70)~\mathrm{D}\,,\\
\bra{e_{A_2}}{}\hat{d}\ket{0}{}&=&(13.58, 3.53 , 1.78 )~\mathrm{D}\,,\\
\bra{e_{B_1}}{}\hat{d}\ket{0}{}&=&(1.50 , 2.60 , -14.20)~\mathrm{D}\,,\\
\bra{e_{B_2}}{}\hat{d}\ket{0}{}&=&(4.98 , -12.51, -3.81 )~\mathrm{D}
\end{eqnarray}
are the dipole moments for qubit A$_1$, A$_2$, B$_1$ and B$_2$ given in \cite{mirkovic_ultrafast_2007}. We take $|\hat{d}_{jk}|\equiv|\hat{d}_{A,jk}|=|\hat{d}_{B,jk}|$, where $j, k= 0-3$. We also take $|\vec{d}_{31}|=|\vec{d}_{10}|$ based on available data which show that the transition strength of the excited state absorption is close to that of the Q$_y$ absorption \cite{dage_density_1999,renger_dissipative_1996,gradinaru_ultrafast_1998}. These physical parameters give $|\vec{d}_{10}|=5 \mathrm{D}$ and other dipole moments are given as ratios in Table \ref{tab:para}.

\section{Light intensity calculation}

\label{sec:appdxLight}

The electric field $E_{0}$ in a laser pulse can be estimated by
\begin{equation}
\frac{\epsilon_{0}E_{0}^2}2= \frac{W}V,
\end{equation}
where $\epsilon_{0}$ and is the permittivity of free space, $W$ is the energy of the pulse and $V=c T A$ is the beam volume given by the speed of light $c$, pulse duration $T$ and beam cross-section $A$. From this, we can define a dimensionless parameter
\begin{equation}
\gamma=\frac{E_0 d T}{\hbar}=\frac{d }{\hbar}\sqrt{ \frac{2WT}{c A\epsilon_{0}}} 
\end{equation}
in terms of known parameters. Pulses used to illuminate photosynthetic systems have typical values of $W= 5\mathrm{nJ}$, $T=10\mathrm{fs}$ and $A=2500\pi \mu \mathrm{m}^2$. We also use  $d=5 \mathrm{D}$ (calculated in \ref{sec:appdxPara}) to estimate a typical value of $\gamma=0.41$.



\end{document}